\pdfoutput=1

\documentclass[aps,twocolumn,amsmath,amssymb,preprintnumbers,floatfix,prb,superscriptaddress,longbibliography]{revtex4-2}%{revtex4}%

\usepackage[utf8]{inputenc}
\usepackage{newtxtext}
\usepackage[upint]{newtxmath}
\usepackage{microtype}
\usepackage{textcomp}
\usepackage{eucal}
\usepackage{bm}
\usepackage{siunitx}
\usepackage{comment}
\usepackage{lipsum}
\usepackage{ulem}

\usepackage{enumerate}
\usepackage{amsfonts}
\usepackage{amsmath}
\usepackage{amssymb}
\usepackage{color}
\usepackage{soul}

\usepackage{graphicx}

\usepackage[colorlinks,allcolors=blue]{hyperref}
\usepackage[capitalize]{cleveref}

\definecolor{DarkRed}{rgb}{0.65,0,0}%
\definecolor{Green}{rgb}{0,0.3,0.3}
\definecolor{Purple}{rgb}{0.3,0,0.65}
\definecolor{Red}{rgb}{1,0,0}
\definecolor{Blue}{rgb}{0,0,0.85}
\definecolor{Magenta}{rgb}{1,0,1}

\newcommand{\be}{\begin{equation}}
\newcommand{\ee}{\end{equation}}

\newcommand{\prlsection}[1]{\textit{#1}.\kern0.05em---\kern0.05em\ignorespaces}

\begin{document}
%\title{Self-consistent Keldysh-Usadel formalism reveals the dominance of crossed Andreev reflection over elastic cotunneling in N/S/N heterostructures}
\title{Crossed Andreev reflection revealed by self-consistent Keldysh-Usadel formalism}
\author{Johanne Bratland Tjernshaugen}
\affiliation{Center for Quantum Spintronics, Department of Physics, Norwegian \\ University of Science and Technology, NO-7491 Trondheim, Norway}
\author{Morten Amundsen}
\affiliation{Center for Quantum Spintronics, Department of Physics, Norwegian \\ University of Science and Technology, NO-7491 Trondheim, Norway}
\author{Jacob Linder}
\affiliation{Center for Quantum Spintronics, Department of Physics, Norwegian \\ University of Science and Technology, NO-7491 Trondheim, Norway}

%%%%%%%%%%%%%%%%
\begin{abstract}
Crossed Andreev reflection (CAR) is a process that creates entanglement between spatially separated electrons and holes. Such entangled pairs have potential applications in quantum information processing, and it is therefore relevant to determine how the probability for CAR can be increased. CAR competes with another non-local process called elastic cotunneling (EC), which does not create entanglement. In conventional normal metal/superconductor/normal metal heterostructures, earlier theoretical work predicted that EC dominates over CAR. Nevertheless, we show numerically that when the Keldysh-Usadel equations are solved self-consistently in the superconductor, CAR can dominate over EC. Self-consistency is necessary both for the conversion from a quasiparticle current to a supercurrent and to describe the spatial variation of the order parameter correctly.
A requirement for the CAR probability to surpass the EC probability is that the inverse proximity effect be small. Otherwise, the subvoltage density of states becomes large and EC is strengthened by quasiparticles flowing through the superconductor. Therefore, CAR becomes dominant in the non-local transport with increasing interface resistance and length of the superconducting region. Our results show that even the simplest possible experimental setup with easily accessible normal metals and superconductors can provide dominant CAR by designing the experimental parameters correctly. We also find that spin-splitting in the superconductor increases the subvoltage density of states, and thus always favors EC over CAR. Finally, we tune the chemical potential in the leads such that transport is governed by electrons of one spin type. This can increase the CAR probability at finite values of the spin-splitting compared to using a spin-degenerate voltage bias, and provides a way to control the spin of the conduction electrons electrically. 
\end{abstract}

\maketitle
 \begin{figure*}[t]
    \centering
    \includegraphics[width=\textwidth]{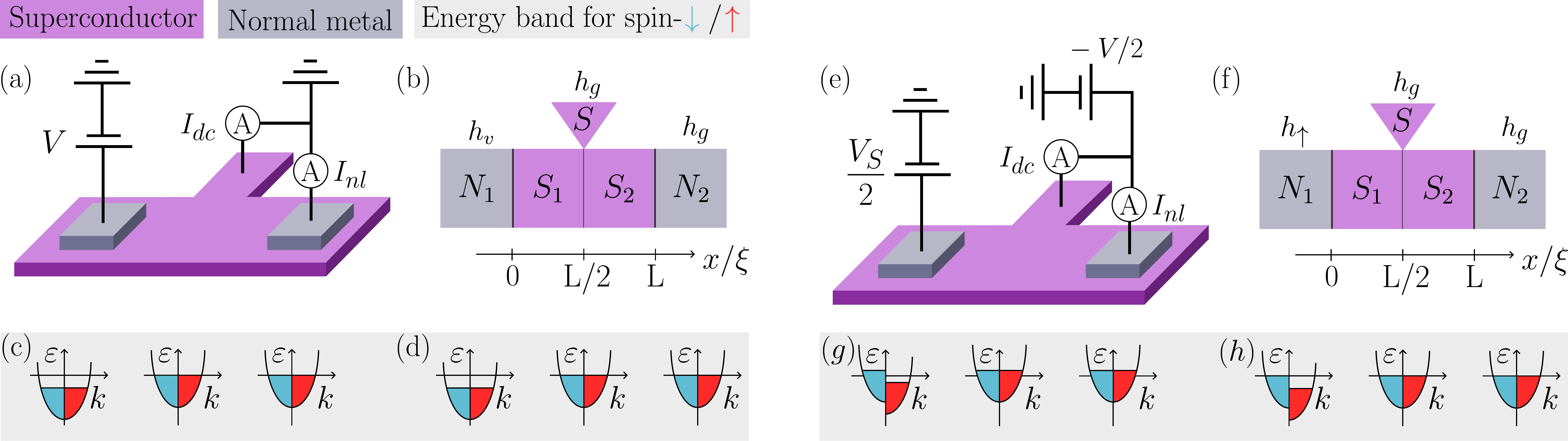}
    \caption{\textbf{(a)} Schematic experimental setup for a superconductor coupled to two normal leads. A drive current $I_{dc}$ flows from the left normal lead and into the grounded superconductor. The non-local current $I_{nl}$ is measured in the second normal metal lead and its sign indicates whether CAR or EC dominates the non-local processes. \textbf{(b)} Theoretical model for the setup in (a). The reservoirs $N_1$ and $N_2$ represent the normal leads, while the superconducting reservoir $S$ represents the grounded part of the superconductor. The distribution function $h_v$ in $N_1$ represents an electrical voltage, and $N_2$ and $S$ are kept at the ground level by setting the distribution functions to $h_g$. For numerical convenience, the superconductor is split into two parts $S_1$ and $S_2$. \textbf{(c, d)} A sketch showing the electron occupation for spin-down (blue) and spin-up (red) in $N_1$, $S$ and $N_2$, respectively. The energy $\varepsilon$ is measured relative to the Fermi energy, and $k$ is the wave number. \textbf{(e)} Schematic experimental setup for spin-dependent transport. A spin voltage $V_S/2$ is applied to $N_1$, while an electrical voltage $-V/2$ shifts the chemical potentials in $S$ and $N_2$. In practice, the spin voltage can also be replaced by a spin accumulation which leaves the band energies intact and instead shifts the chemical potentials of the two spin species oppositely, for instance via spin pumping. \textbf{(f)} Theoretical model for the setup in (e). We redefine the ground level to align with $S$ and $N_2$ since only differences in chemical potentials matter. \textbf{(g, h)} A sketch of the electron occupation in (e) and (f), respectively. 
    }
    \label{fig:car-setup}
\end{figure*}

%%%%%%%%%%%%%%%%
\section{Introduction} 
Quantum information processing, such as quantum cryptography and teleportation, offers exciting possibilities for future computation and communication technologies \cite{wendin_quantum_2017, ladd_quantum_2010, pirandola_advances_2020, pirandola_advances_2015}. One of the stepping stones toward such technologies is to create entanglement between particles that are separated in space \cite{horodecki_quantum_2009}. Superconductors provide a natural platform for creating entanglement since the Cooper pair consists of two entangled electrons. Crossed Andreev reflection (CAR) is a process that exploits the correlations in the Cooper pair to create entanglement between spatially separated electrons and holes.  

CAR can be explained in a superconducting hybrid system where a Bardeen-Cooper-Schrieffer (BCS) \cite{bardeen_theory_1957} superconductor is coupled to three different leads, as sketched in Fig. \ref{fig:car-setup}(a). We illustrate the concept here with two normal leads $N_1$ and $N_2$ and one superconducting lead $S$ as in Fig. \ref{fig:car-setup}(b). Two of the leads, $N_2$ and $S$, are grounded, while the chemical potential is shifted to $eV<0$ in the third lead $N_1$. The electron occupation in the different leads is illustrated in Fig. \ref{fig:car-setup}(c)-(d). Here, $V < \Delta_0/|e|$ is the voltage bias, $e<0$ is the electronic charge, and $\Delta_0$ is the bulk superconducting gap at zero temperature. This causes a drive current to flow into $N_1$ from the closest grounded lead $S$, or equivalently it causes electrons to flow from $N_1$ to $S$. However, electrons in $N_1$ cannot enter the superconductor as long-lived quasiparticles due to the gap in the density of states (DOS). Instead, the incoming electrons in $N_1$ might pair up with electrons of opposite spin from $N_1$ or $N_2$, and together they can enter the superconductor as a Cooper pair. These processes are called Andreev reflection (AR) \cite{andreev_thermal_1964} and crossed Andreev reflection, respectively. AR causes a hole to be reflected into $N_1$, while CAR causes a hole to propagate away from the superconductor in $N_2$. This hole is entangled with the incoming electron. Another possibility is that the incoming electron is directly transmitted from $N_1$ to $N_2$ via for example charge imbalance, which is the conversion from a resistive current to a supercurrent, or via a virtual state in the superconductor. We call this elastic cotunneling (EC). EC competes with CAR, and since EC does not induce entanglement, we want to increase the CAR signal relative to EC.  

The sign of the non-local current over the $S_2/N_2$ interface determines whether CAR or EC dominates the non-local transport. We define the drive current $I_{dc}$ as flowing from $N_1$ to $S$, and the non-local current $I_{nl}$ as flowing from $S$ to $N_2$. The currents can be represented in terms of the voltages $V$ and $V_{nl}$ on the normal leads $N_1$ and $N_2$, respectively, as \cite{falci_correlated_2001, cadden-zimansky_charge_2007}
\begin{equation}\label{eq:relate currents to voltages}
    \begin{pmatrix}
        I_{dc} \\ I_{nl}
    \end{pmatrix} = 
    \begin{pmatrix}
        G_{A} + G_{C}+G_{E} & G_{C}-G_{E} \\ -G_{C}+G_{E} & -G_{A} - G_{C}-G_{E}
    \end{pmatrix}
    \begin{pmatrix}
        V \\ V_{nl}
    \end{pmatrix}.
\end{equation}
Here, we assumed that the left and right interfaces are identical. $G_A$ is the Andreev conductance, $G_C$ is the crossed Andreev conductance, and $G_E$ is the EC conductance. If we ground the right lead by setting $V_{nl}=0$, we find $I_{nl} = (G_E-G_C)V$. Thus, the non-local current is positive if $G_E > G_C$, meaning that EC dominates the non-local transport. The non-local current is negative if $G_C > G_E$, meaning that CAR dominates the non-local transport. On the other hand, if we invert Eq. \eqref{eq:relate currents to voltages} to write the voltages in terms of the currents, we have 
\begin{equation}
   \begin{pmatrix}
        V \\ V_{nl}
    \end{pmatrix}=
    \frac{1}{a}
     \begin{pmatrix}
        -G_{A} - G_{C}-G_{E} & -G_{C}+G_{E} \\
        G_{C}-G_{E} &  G_{A} + G_{C}+G_{E}
    \end{pmatrix}
    \begin{pmatrix}
        I_{dc} \\ I_{nl}
    \end{pmatrix},
\end{equation}
with the determinant $a<0$ since the conductances are positive. In an open circuit where $N_2$ is not grounded, a voltage builds up until $I_{nl}=0$. In this case, $V_{nl} = -(G_C-G_E)I_{dc}/|a|$. When the drive current is positive, we see that $V_{nl}>0$ when EC dominates and $V_{nl}<0$ when CAR dominates. Numerically, it is easier to calculate the non-local current than the non-local voltage.

A great collection of literature, both theoretical and experimental, exists on how the CAR signal can be enhanced relative to the EC signal. Using ferromagnetic leads \cite{falci_correlated_2001, yamashita_crossed_2003, deutscher_coupling_2000, morten_elementary_2008} instead of normal leads enhances CAR in an antiparallel setup and enhances EC in the parallel setup. The probabilities of CAR and EC can be tuned by changing the gate voltage of an intermediate normal region \cite{soori_tunable_2022}. However, ferromagnets cause stray fields that could act disruptively to neighboring elements in a device architecture. Other proposals include using spin-polarized interfaces \cite{kalenkov_crossed_2007}, antiferromagnetic leads \cite{jakobsen_electrically_2021}, altermagnetic leads \cite{das_crossed_2024}, an ac voltage bias \cite{golubev_non-local_2009}, and graphene \cite{cayssol_crossed_2008, veldhorst_nonlocal_2010, tan_cooper_2015,borzenets_high_2016, pandey_ballistic_2021}. In the simplest system that displays CAR, namely a conventional superconductor with normal leads, the CAR signal is theoretically predicted \cite{cadden-zimansky_nonlocal_2006, brinkman_crossed_2006, morten_circuit_2006, morten_full_2008, kalenkov_nonlocal_2007, feinberg_andreev_2003, melin_self-consistent_2009, floser_absence_2013, freyn_positive_2010, bergeret_nonlocal_2009} to be smaller or equal in magnitude compared to the EC signal. On the other hand, experiments \cite{russo_experimental_2005, kleine_contact_2009} show that CAR may dominate the non-local signal. Using normal leads, such as copper, does not require fine-tuning of the electronic structure or other parameters, or any rare materials, and this is an advantage compared to many of the previous works.
 
CAR has previously been studied in the quasiclassical Keldysh-Usadel formalism in a normal/superconductor/normal (NSN) heterostructure \cite{brinkman_crossed_2006, bergeret_nonlocal_2009}, but not in a fully self-consistent manner. Specifically, the spatial variation of the gap and the resistive and dissipationless currents in the superconductor have not been considered in these works. However, accounting for these properties through full self-consistency is important because otherwise the conversion from resistive to supercurrent is not captured, nor the spatial modulation of the superconducting gap, which both affect the local density of states and the probabilities for EC and CAR. This becomes increasingly important the longer the superconductor is, unlike a short superconductor smaller than the coherence length \cite{bergeret_nonlocal_2009}. We consider the system shown in Fig. \ref{fig:car-setup}(a)-(b) and find numerically that CAR can dominate over EC when the superconducting order parameter is determined self-consistently. This underlines the importance of self-consistency when solving the Usadel equation.

We find that the subvoltage density of states is a crucial factor in determining whether CAR or EC dominates. In a bulk superconductor, the subgap density of states is zero. However, when a short superconductor is proximitized with a non-superconductor, the inverse proximity effect causes the subgap density of states to be suppressed, but not exactly zero. This means that there are available quasiparticle states also for small energies, and therefore electrons can travel through the superconductor as quasiparticles. Hence, if the subvoltage density of states is large, EC will dominate. Additionally, a resistive current is converted to a supercurrent over the length scale of the coherence length. CAR gives rise to transport by supercurrents and is also strengthened by the presence of a supercurrent \cite{chen_long-range_2015}, and thus CAR is suppressed in short superconductors where the supercurrent is negligible.

Furthermore, we determine how the presence of a spin-splitting $m$ in the superconductor affects the ratio of the CAR signal. Spin-splitting in a thin film superconductor can be achieved by growing a ferromagnetic insulator underneath it or by applying an in-plane magnetic field. The motivation behind introducing spin-splitting is that the combination of magnetism and superconductivity may enhance existing phenomena and give rise to new phenomena, such as giant thermoelectric effects, huge magnetoresistance effects, and very long spin diffusion lengths \cite{linder_superconducting_2015, eschrig_spin-polarized_2011, bergeret_colloquium_2018}. Spin-splitting will affect EC in a different way than CAR because CAR involves electrons with opposite spins while EC involves one electron with one spin. We find that spin-splitting increases the subgap density of states for two reasons. First, it increases because the peaks in the density of states for spin-up and spin-down quasiparticles are shifted relative to each other. This enables quasiparticle states to be available for transport in the superconductor at smaller bias voltages than without spin-splitting. Second, spin-splitting causes the gap to decrease, and then the DOS at low energies increases. Therefore, spin-splitting always favors EC over CAR.

Nevertheless, if we filter out one of the spin bands inside the superconductor, shifting the peaks in the DOS due to spin-splitting does not increase the relevant density of states. If the spin-splitting does not increase the low-energy DOS too much, we demonstrate that a spin-dependent voltage allows for CAR in regimes where an electrical voltage would give EC. Spin-filtering can be achieved by using ferromagnetic leads or spin-filtering at the interfaces between the superconductor and the leads \cite{falci_correlated_2001, beckmann_negative_2007, kalenkov_crossed_2007}. We propose a different means of achieving spin-dependent transport by tuning the chemical potentials in the leads, as illustrated in Fig. \ref{fig:car-setup}(h).
In effect, the chemical potential for spin-up electrons changes in the different leads, while for spin-down electrons the chemical potential is the same in the entire system. Therefore, we only get transport of spin-up electrons and holes representing missing spin-up electrons. In practice, this could be achieved by applying a spin accumulation $V_S/2$ in $N_1$, as shown in Fig. \ref{fig:car-setup}(e). This shifts the chemical potential to $|e|V_S/2$ for spin-down electrons and to $-|e|V_S/2$ for spin-up electrons. If at the same time we apply an electrical voltage $-V/2 = -V_S/2$ to $S$ and $N_2$, the chemical potential for spin-down electrons will be $|e|V_S/2$ in all leads while the chemical potential will vary between $\pm|e|V_S/2$ for spin-up electrons, see Fig. \ref{fig:car-setup}(g). We find that we cannot increase the probability of CAR by spin-dependent transport. The only exception is when the voltage and the spin-splitting are both large, but then CAR could also be restored by removing the spin-splitting.

%%%%%%%%%%%%%%%%%
\section{Model} 

The quasiclassical, non-equilibrium Keldysh Green function theory \cite{belzig_quasiclassical_1999, bergeret_colloquium_2018, chandrasekhar_superconductivity_2008} 
is a powerful formalism for calculating observables in superconducting hybrid systems. The $8\times 8$ Green function $\check{G}$ in Keldysh$\otimes$Nambu$\otimes$spin space is defined in terms of the spinor
\begin{equation}
    \psi(\boldsymbol{r},t) = \begin{pmatrix}
        \psi_{\uparrow} (\boldsymbol{r},t) & \psi_{\downarrow}(\boldsymbol{r},t) & \psi_{\uparrow}^{\dagger}(\boldsymbol{r},t) & \psi_{\downarrow}^{\dagger}(\boldsymbol{r},t)
    \end{pmatrix}^T,
\end{equation}
where $\psi_{\sigma}^{\dagger}$ and $\psi_{\sigma}$ are the electronic creation and annihilation operators, respectively. The Green function has the matrix structure
\begin{equation}
    \check{G}=\begin{pmatrix}
        \hat{G}^R & \hat{G}^K \\ 0 & \hat{G}^A
    \end{pmatrix},
\end{equation}
where the retarded Green function $\hat{G}^R$, the advanced Green function $\hat{G}^A$ and the Keldysh Green function $\hat{G}^K$ are defined by 
\begin{align*}
    \hat{G}^R(\boldsymbol{r},t,\boldsymbol{r}',t') &= -i\theta(t-t')\hat{\rho}_4\langle \{ \psi(\boldsymbol{r},t), \psi^{\dagger}(\boldsymbol{r}',t') \} \rangle, \\   \hat{G}^A(\boldsymbol{r},t,\boldsymbol{r}',t') &= +i\theta(t'-t)\hat{\rho}_4\langle \{ \psi(\boldsymbol{r},t), \psi^{\dagger}(\boldsymbol{r}',t') \} \rangle, \\ \hat{G}^K(\boldsymbol{r},t,\boldsymbol{r}',t') &= -i\hat{\rho}_4\langle [ \psi(\boldsymbol{r},t), \psi^{\dagger}(\boldsymbol{r}',t') ] \rangle.
\end{align*}
The matrix $\hat{\rho}_4=\text{diag}(1,1,-1,-1)$ is a part of a set of matrices $\{\hat{\rho}_n\}$ that span the block-diagonal Nambu$\otimes$spin space.  They are defined as $\hat{\rho}_n = \tau_0 \otimes \sigma_n$ for $n\in \{0,1,2,3\}$ and $\hat{\rho}_n = \tau_3 \otimes \sigma_{n-4}$ for $n \in \{4,5,6,7\}$. The Green function $\check{G}$ is exact and can in principle be determined from the Gorkov equation \cite{gorkov_energy_1958}, but in practice this is prohibitively difficult. The problem is simplified in the quasiclassical limit, where all energy scales in the system are much smaller than the Fermi energy. In such systems, $\check{G}$ is strongly peaked near the Fermi momentum $\boldsymbol{p}_F$ \cite{belzig_quasiclassical_1999, chandrasekhar_superconductivity_2008}. The quasiclassical Green function $\check{\Gamma}$ is defined by constricting $\check{G}$ to the Fermi surface, or equivalently by the integral
\begin{equation}
    \check{\Gamma}\left(\varepsilon, \frac{\boldsymbol{p}_F}{|\boldsymbol{p}_F|}\right) = \frac{i}{\pi} \int_{-\omega_c}^{\omega_c} \check{G}(\varepsilon, \boldsymbol{p}) \text{d}\epsilon_{\boldsymbol{p}},  
\end{equation}
where $\epsilon_{\boldsymbol{p}}= {(|\boldsymbol{p}|^2-|\boldsymbol{p}_F|^2)}/{2m}$. The energy $\varepsilon$ is measured relative to the Fermi energy. The quasiclassical Green function obeys the Eilenberger equation \cite{eilenberger_transformation_1968}, which can be reduced to a diffusion-like equation in the dirty limit. In the dirty limit one assumes that the impurity concentration is high, such that the elastic mean free path is much smaller than any other length scale in the system except the Fermi wavelength. In this limit, the quasiclassical Green function should depend weakly on the transport direction $\boldsymbol{p}_F/|\boldsymbol{p}_F|$. The equation can then be expressed in terms of the isotropic, quasiclassical Green function $\check{g}$, which we from here term the Green function:
\begin{equation}
    {\frac{\partial}{\partial (x/\xi)}\left(\check{g} \frac{\partial\check{g}}{\partial (x/\xi)}\right) }= -i[\varepsilon\hat{\rho}_4+\hat{\Delta}+\hat{M},\check{g}]/\Delta_0.
\end{equation}
This is the Usadel equation \cite{usadel_generalized_1970}. The self-energy $\hat{M} = \text{diag}(m\sigma_z, m \sigma_z)$ describes spin-splitting of strength $m$ in the $z$-direction. The superconducting order parameter, defined as $\Delta(\boldsymbol{r}) = -\lambda \langle \psi_{\downarrow}(\boldsymbol{r}) \psi_{\uparrow}(\boldsymbol{r}) \rangle$, enters the equation in the self-energy $\hat{\Delta} = \text{antidiag}(\Delta, -\Delta, \Delta^*, -\Delta^*)$. The time coordinate was omitted because we consider time-independent systems, while importantly we keep the spatial coordinate in the order parameter.
Inelastic scattering in the Usadel equation is modeled using the Dynes approximation $\varepsilon \rightarrow \varepsilon + i\delta$. This is a good approximation for spectral properties, but it does not model energy loss in the sense that the occupation of holes and electrons at any energy is unaffected by inelastic scattering. Thus, we are assuming that the superconductor is much shorter than the inelastic scattering length.

The Green function allows us to calculate observables such as the charge current $I$,
\begin{equation} \label{eq: charge current}
    I = -I_0 \int_{0}^{\infty} \text{Re Tr}\left[\hat{\rho}_4\left(\check{g} \frac{\partial \check{g}}{\partial (x/\xi)} \right)^K \right]\text{d}\left(\frac{\varepsilon}{\Delta_0}\right),
\end{equation}
or the superconducting order parameter,
\begin{equation}\label{eq:gap equation}
    \frac{\Delta}{\Delta_0} = -\frac{N_0\lambda}{4}\int_{-\omega_c}^{\omega_c} \hat{g}^K_{23}\left(\frac{\varepsilon}{\Delta_0}\right)\text{d}\left(\frac{\varepsilon}{\Delta_0}\right).
\end{equation}
The prefactor $I_0=A|e|N_0\Delta_0^2\xi/8\hbar$, with $A$ being the interfacial contact area, $N_0$ the normal-state density of states at the Fermi level, and $\xi$ the superconducting coherence length, is typically of the order $I_0/A \propto 10^{9} \text{A/m²}$. For eq. \eqref{eq:gap equation} to be consistent, the coupling constant $\lambda$ is related to the cutoff energy by $\omega_c = \text{cosh}(1/N_0\lambda)$. The order parameter can be calculated from the solution $\check{g}$ of the Usadel equation, but it also enters the equation as a fixed self-energy. Therefore, the Usadel equation must be solved self-consistently for the order parameter using fixed-point iterations.

The Keldysh Green function is given by 
\begin{align}
\hat{g}^K = \hat{g}^R\hat{h}- \hat{h}\hat{g}^A,
\end{align}
where the distribution function $\hat{h} = h_n \hat{\rho}_n$ describes the occupancy of electron and hole states. The nonequilibrium modes $h_n$ are associated with different electronic degrees of freedom. For example, the charge mode $h_4$ is generated by electric currents and it is an imbalance of the occupation of electron and hole states. This can be seen by rewriting $\hat{h}$ in terms of the electron and hole occupation probabilities $f_{e,\sigma}$ and  $f_{h,\sigma}$ \cite{distributionfnc-form},
\begin{equation}\label{eq: disfnc in terms of occupation}
    \hat{h} = \hat{1} - 2\cdot \text{diag}(f_{e,\uparrow}, f_{e,\downarrow},f_{h,\downarrow}, f_{h,\uparrow}),
\end{equation}
where we define a hole with spin $\sigma$ as a missing electron with spin $-\sigma$. Thus,
\begin{equation}
    h_4 = \frac{1}{4}\text{Tr}\left( \hat{\rho}_4 \hat{h} \right) = \frac{1}{2}(f_{e,\uparrow}+ f_{e,\downarrow} -f_{h,\downarrow}- f_{h,\uparrow}).
\end{equation}
In equilibrium, the electron and hole occupation probabilities are given by the Fermi-Dirac distribution $f(\varepsilon)$. In the presence of a spin-dependent voltage $V_{\sigma}$, they are given by $f_{e,\sigma} = f(\varepsilon + |e|V_{\sigma})$ and $f_{h,\sigma} = f(\varepsilon + |e|V_{-\sigma})$. We define the ground level to be at $V=0$, and thus the distribution function $\hat{h}_g$ at the ground level is 
\begin{equation}\label{eq:disfnc for ground}
    \hat{h}_g = \text{diag}(t_0,t_0,t_0,t_0),
\end{equation}
where $t_{0} = \tanh{[\varepsilon/2T]}$, and $T=0.01T_c$ is the temperature we use in our simulations.  An applied standard electrical voltage is modeled by the distribution function $\hat{h}$ as
\begin{equation}\label{eq:disfnc for a standard voltage}
    \hat{h}_v=\text{diag}(t_+, t_+, t_-, t_-),
\end{equation}
where $t_{\pm} = \tanh[(\varepsilon \pm |e|V)/2T]$.
The spin-up equivalent $\hat{h}_{\uparrow}$ describing transport of spin-up electrons is 
\begin{equation}\label{eq:disfnc for a spin up voltage}
    \hat{h}_{\uparrow} =\text{diag}(t_+, t_0, t_-, t_0).
\end{equation}
This is derived from eq. \eqref{eq: disfnc in terms of occupation} by setting $V_{\downarrow}=0$.
We will write that we use an electrical voltage when the distribution function in $N_1$ is given by eq. \eqref{eq:disfnc for a standard voltage}, as in Fig. \ref{fig:car-setup}(b), and that we use a spin-up voltage $V_{\uparrow}$ when it is given by eq. \eqref{eq:disfnc for a spin up voltage}, as in Fig. \ref{fig:car-setup}(f).

The retarded Green function and the corresponding Usadel equation were Riccati parametrized \cite{schopohl_quasiparticle_1995, konstandin_superconducting_2005},
\begin{equation}
    \hat{g}^R = \begin{pmatrix}
        N&0\\0&-\Tilde{N}
    \end{pmatrix} \begin{pmatrix}
        1+\gamma\Tilde{\gamma} & 2\gamma \\ 2\Tilde{\gamma} & 1+\Tilde{\gamma}\gamma
    \end{pmatrix}.
\end{equation}
This parametrization is convenient for numerical calculations because it is single-valued and the parameters are bound to the interval $[0,1]$. The kinetic equations for the distribution function were parametrized using the $\hat{\rho}_n$--matrices defined earlier, similarly to Ref. \cite{ouassou_voltage-induced_2018}.

The Usadel equation is accompanied by boundary conditions. At the normal interfaces, we use the tunneling Kuprianov-Lukichev boundary conditions \cite{kuprianov_influence_1988}, 
\begin{equation}
    \check{g} \frac{\partial\check{g}}{\partial (x/\xi)} = \frac{\pm 1}{2Lr}[\underline{\check{g}},\check{g}],
\end{equation}
when the normal metals are to the left and right of the superconductor, respectively. Here, $\check{g}$ is the Green function at the interface while $\underline{\check{g}}$ is the Green function in the reservoir. We normalize the length $l$ of the superconductor on the coherence length so that the dimensionless measure $L$ of the length of the superconductor is $L=l/\xi$. The interfaces are characterized by the interface parameter $r = r_B/r_0$. A high interface parameter corresponds to a high barrier resistance $r_B$ compared to the bulk resistance $r_0$, and conversely a low interface transparency. \
The Kuprianov-Lukichev boundary conditions can be justified by comparing the expression for the quasiclassical current with a current calculated from a tunneling Hamiltonian \cite{bergeret_electronic_2012, linder_quasiclassical_2022}. The original derivation was done by Kuprianov and Lukichev \cite{kuprianov_influence_1988}, while the boundary conditions were more thoroughly derived and also generalized to magnetic interfaces by Ref. \cite{cottet_spin-dependent_2009}.

\begin{figure}
    \centering
    \includegraphics[width=\linewidth]{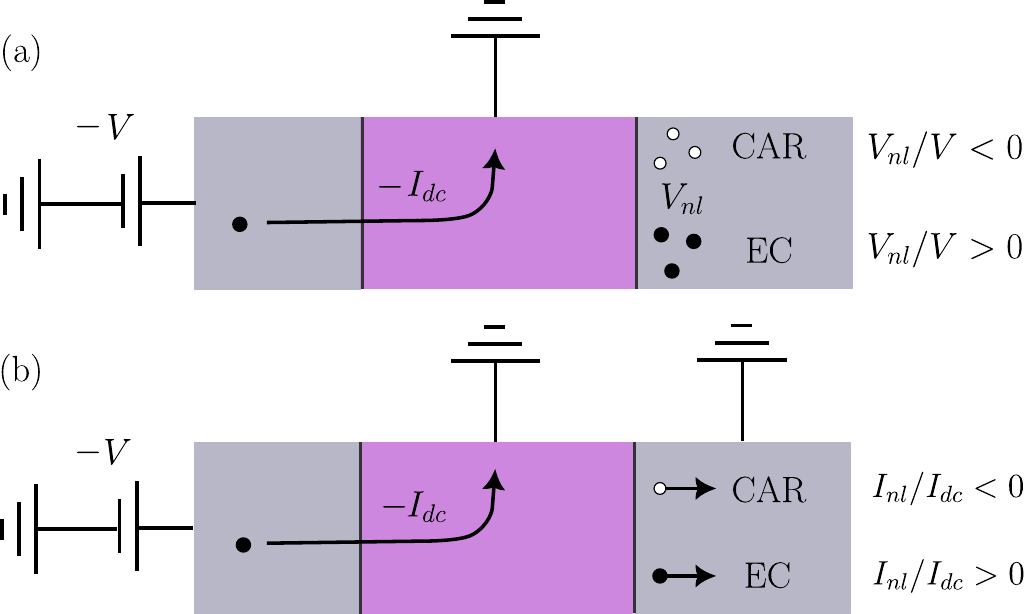}
    \caption{Illustration of the difference between the second normal lead being \textbf{(a)} open-ended or \textbf{(b)} grounded. The minus signs in the bias voltage and drive current are chosen to illustrate the CAR and EC processes in terms of an incoming electron. In the open-ended setup, charge accumulates in the second normal lead. The charge has opposite sign for the CAR and EC processes because CAR causes an accumulation of holes, while EC causes an accumulation of electrons. When the second normal lead is grounded, the non-local currents induced by CAR and EC have opposite signs. }
    \label{fig:open-ended vs grounded}
\end{figure}

The system that we consider is a superconductor coupled to two normal leads $N_1$ and $N_2$,  as shown in Fig. \ref{fig:car-setup}(a). In an experiment, the superconductor is typically connected to ground whereas a voltage difference is applied between one of the normal leads and the superconductor. This induces either a non-local voltage or current in the second normal lead, depending on whether the second normal lead is open-ended or grounded, respectively. Figure \ref{fig:open-ended vs grounded} shows the difference between the open-ended and grounded setup. 

In the case where this lead is open-ended, charge will accumulate until the net current into the second normal lead is zero. The accumulated charge is measured as a non-local voltage. It is called non-local because it appears in a part of the system where no current is flowing. Therefore, in order for a non-local voltage to exist, it is necessary to drain currents from the superconductor. In a two-terminal setup, in effect a superconductor coupled to two normal leads, the currents in the two leads are identical due to current conservation. Therefore, it is impossible to induce a non-local voltage by driving a current through a distant part of the system. It is necessary to have at least three terminals in order to model a non-local voltage. 

In some experiments, the second normal lead is instead grounded, and one measures a current through this lead. We refer to this current as the non-local current, since this is the current that would build up a non-local voltage for an open-ended circuit. This scenario is shown in Fig. \ref{fig:car-setup}(a) and this is the case we are considering for the most part in this paper.

Importantly, the sign of the current flowing in this lead contains the same information as the sign of voltage in the open-ended geometry. The reason for this is shown in Fig. \ref{fig:open-ended vs grounded}.
If the current flows into the grounded second normal lead due to EC, the sign of the non-local voltage will be the same as in the first lead. If the current instead flows out of the grounded second normal lead due to CAR, a non-local voltage of the opposite sign must be induced. Therefore, instead of looking at the sign of the induced voltage in an open-ended system, one can look at the sign of the current into the grounded second normal lead. 
It is simpler and faster to numerically calculate the current than the non-local voltage, and it is also viable to measure this current experimentally \cite{zhang_next_2019}. We will verify the correspondence between the sign of the current in the grounded setup and the voltage in the open-ended setup with numerical simulations later in this paper.

Finally, we comment on the role of the superconducting reservoir in our setup. We want to model a current flowing from a normal lead into the superconducting bar and into ground. One could then use a 2D geometry shaped like a T [as shown in Fig. \ref{fig:car-setup}(a)] and solve self-consistently throughout the entire superconductor. However, solving the non-equilibrium Usadel equation self-consistently in a 2D geometry is very time consuming. Therefore, we replace the part of the superconductor that is connected to ground with a tunneling interface into a grounded superconducting reservoir. The tunneling interface is assumed to be quite transparent since it represents an actual part of the superconductor. Since currents are allowed to flow into the superconducting reservoir, the current is not constant in the remaining part of the superconductor due to current conservation. We therefore solve the Usadel equation self-consistently on each side of the connection to the superconducting reservoir.
As illustrated in Fig. \ref{fig:car-setup}(b), our model then consists of the first normal lead $N_1$ connected to one part of the superconductor $S_1$. The second normal lead $N_2$ is connected to another part of the superconductor $S_2$, and $S_1$ and $S_2$ are connected with boundary conditions demanding the Green function to be continuous. At the connection between $S_1$ and $S_2$, a current can flow into the grounded superconducting reservoir $S$.

The boundary conditions we use to connect the different superconducting regions are the following \cite{titov_thermopower_2008}. The Green function is demanded to be continuous at the $S_1/S_2$ interface through the boundary condition
\begin{equation}\label{eq:bc_continuity}
    \check{g}_1(L/2) = \check{g}_2(L/2).
\end{equation}
Here, $\check{g}_1$ and $\check{g}_2$ are the Green functions in $S_1$ and $S_2$, respectively. At the $S_1/S/S_2$ intersection, we demand the current to be conserved,
\begin{equation}\label{eq:bc_current_conservation}
     \left[\check{g}_{2}  \frac{\partial \check{g}_{2}}{\partial (x/\xi)} -  \check{g}_{1}  \frac{\partial\check{g}_{1}}{\partial (x/\xi)} \right]\left(\frac{L}{2}\right) = \frac{1}{2Lr_S} [\check{g}_S, \check{g}_{1}\left(\frac{L}{2}\right)].
\end{equation}
The left hand side of this equation is the difference between the currents in $S_1$ and $S_2$, and the right hand side is the current that tunnels into the superconducting reservoir. The interface parameter  $r_S=1$ means that the barrier resistance is equal to the normal state bulk resistance.  The retarded part of the bulk Green function $\check{g}_S$ in $S$ has the Riccati matrices
\begin{equation}
    \gamma_S = \begin{pmatrix}
        0 & \frac{s_{\uparrow}}{1+c_{\uparrow}} \\ \frac{s_{\downarrow}}{1+c_{\downarrow}} & 0
    \end{pmatrix} \qquad \Tilde{\gamma}_S = \begin{pmatrix}
        0 & \frac{s_{\downarrow}}{1+c_{\downarrow}} \\ \frac{s_{\uparrow}}{1+c_{\uparrow}} & 0
    \end{pmatrix}
\end{equation}
where $s_{\sigma} = \sinh(\vartheta_{\sigma})$, $c_{\sigma} = \cosh(\vartheta_{\sigma})$, and $\vartheta_{\sigma} = \text{atanh}(\sigma\Delta_0 / (\sigma m + \varepsilon + i\delta))$ \cite{linder_strong_2015}.

We treat the leads $N_1$, $N_2$ and $S$ as reservoirs. Such an approach was also chosen in Ref. \cite{melin_self-consistent_2009}. The leads could in principle be treated self-consistently by introducing intermediate normal or superconducting layers at the interfaces, but it is unlikely that this would change the results qualitatively. For instance, Ref. \cite{seja_quasiclassical_2021} took the proximity effect in a normal lead into account in a voltage-biased NSN structure and found that self-consistency in the normal leads simply leads to an effectively smaller voltage across the superconducting region, since part of the voltage drop can now occur in the N parts. The magnitude of the superconducting gap may be slightly affected near the interface by treating the normal leads self consistently.

Numerically, the Usadel equation is solved by fixed-point iterations by first guessing a value $\Delta=\Delta_0$ for the order parameters in $S_1$ and $S_2$. Then the retarded equations are solved, the kinetic equations are solved, and the order parameters are calculated from the Green functions in $S_1$ and $S_2$. This is repeated until the absolute changes in the real and imaginary parts of the order parameters fall below the threshold value $10^{-6}$. In principle, one could solve the equations first in $S_1$ using the boundary condition given in eq. \eqref{eq:bc_continuity} and then in $S_2$ using the other boundary condition given in eq. \eqref{eq:bc_current_conservation}. However, we found that the numerical calculations were more stable and accurate when we solved the equations in $S_1$ and $S_2$ simultaneously, thereby ensuring that the boundary conditions were always satisfied.

In the setup shown in Fig. \ref{fig:car-setup}, $N_2$ is grounded and we calculate the non-local current. This is possible to achieve experimentally \cite{zhang_next_2019}, but it is usually simpler to measure the induced non-local voltage for which the non-local current disappears \cite{cadden-zimansky_charge_2007}. This can be done numerically by solving the equation $I_{nl}(V_{nl}) = 0$.

Finally, we note that the numerical values for the spin-splitting and the voltage bias should be carefully chosen to avoid bistability. Bistability means that the order parameter converges to different values depending on the initial guess, which all correspond to local minima in the free energy.
For example, for the interval $\Delta_0/2 < m < \Delta_0$  there exists both a normal and a superconducting solution to the Usadel equation in equilibrium, which are both numerically stable \cite{ouassou_voltage-induced_2018}. Physically, one of these states would be metastable. The ground state switches from superconducting to normal at the Chandrasekhar-Clogston limit $m_c=\Delta_0/\sqrt{2}$ \cite{clogston_upper_1962, chandrasekhar_note_1962}. An out-of-equilibrium superconductor is to a great extent equivalent to a spin-split equilibrium superconductor \cite{moor_inhomogeneous_2009}, so in the presence of a voltage bias, the ground state switches from superconducting to normal at $|e|V_c = \Delta_0/\sqrt{2}$.
Nevertheless, it is hard to calculate free energies in the Keldysh-Usadel formalism and thus numerically determine which state is the physical ground state. Therefore, we choose combinations $m \leq \Delta_0/2$ and $|e|V\leq\Delta_0/2$ for which the superconducting state is the ground state \cite{ouassou_voltage-induced_2018}. In this parameter regime, it is reasonable to approximate the grounded part of the superconductor with a reservoir. For high voltages and spin-splitting fields, the superconducting reservoir should be replaced with a normal reservoir. This means that our model is not valid for the high voltages considered in Ref. \cite{bergeret_nonlocal_2009}.

%%%%%%%%%%%%%%%%%%%%%%%%%%%%%%%
\section{Results and Discussion}
\begin{figure}
    \centering
    \includegraphics[width=\linewidth]{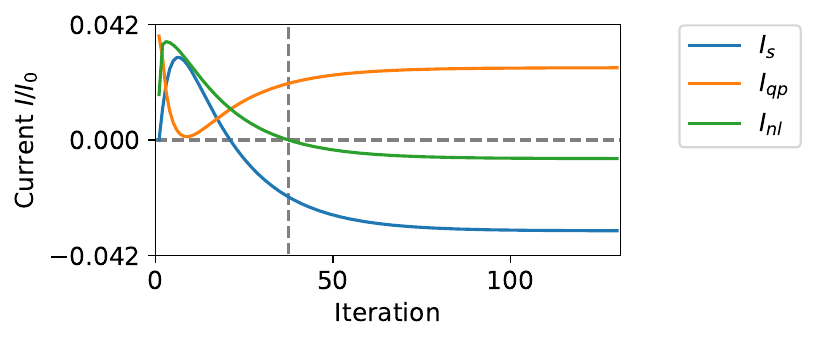}
    \caption{The evolution of various currents as functions of fixed point iterations. The supercurrent $I_s$ and the quasiparticle current $I_{qp}$ are calculated in the middle of the superconductor at the $S_2$ side, i.e. at $x/\xi=(L/2)^+$. The non-local current $I_{nl}$ changes sign at iteration number $38$. The parameters used are $r=2.5$, $L=2.125$, $m=0$ and $|e|V=0.3\Delta_0$.}
    \label{fig:iterations}
\end{figure}

% The importance of self consistency
Fig. \ref{fig:iterations} shows how the non-local current develops over the fixed point iterations for a given choice of length, interface transparency, and applied voltage. In the beginning, the non-local current acquires an increasingly positive value, indicating that EC dominates the non-local transport. After some more iterations, the non-local current starts decreasing. At some point the non-local current switches sign, and then it stabilizes to a negative value. Therefore, CAR dominates the non-local transport. This illustrates that self-consistency in the Usadel equation can reveal physics that cannot be seen unless the equation is solved self-consistently. Fig. \ref{fig:iterations} also shows that the supercurrent and the quasiparticle current in the middle of the superconductor vary much over the first $50$ iterations. Note that the number of iterations needed for the currents to stabilize depends on parameter choices such as the length and the bias voltage. Self-consistency assures that the supercurrent and the quasiparticle currents converge to their true values, which again assures that the non-local current converges to its true value. 

\begin{figure}
    \centering
    \includegraphics[width=\linewidth]{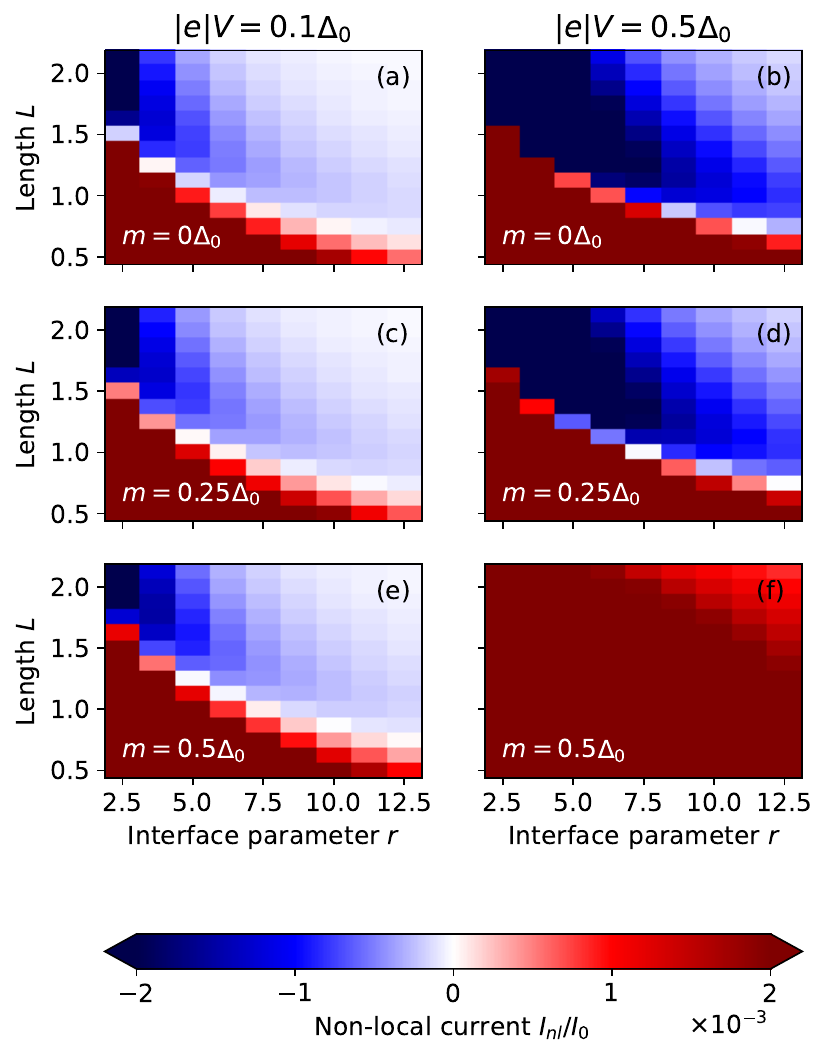}
    \caption{The non-local current $I_{nl}/I_0$ in the $(r,L)$-plane for different combinations of the spin-splitting $m$ and the voltage bias $V$. The left column has voltage bias $|e|V = 0.1\Delta_0$, while the right column has $|e|V = 0.5\Delta_0$. EC dominates the non-local transport when $I_{nl} >0$, colored red in the plot. In the blue regions, $I_{nl}<0$ and CAR dominates. Note that the color minimum and maximum represent currents $|I_{nl}|/I_0 > 2\cdot10^{-3}$.}
    \label{fig:current-rLplane-standard-voltage}
\end{figure}

\begin{figure}
    \centering
    \includegraphics[width=\linewidth]{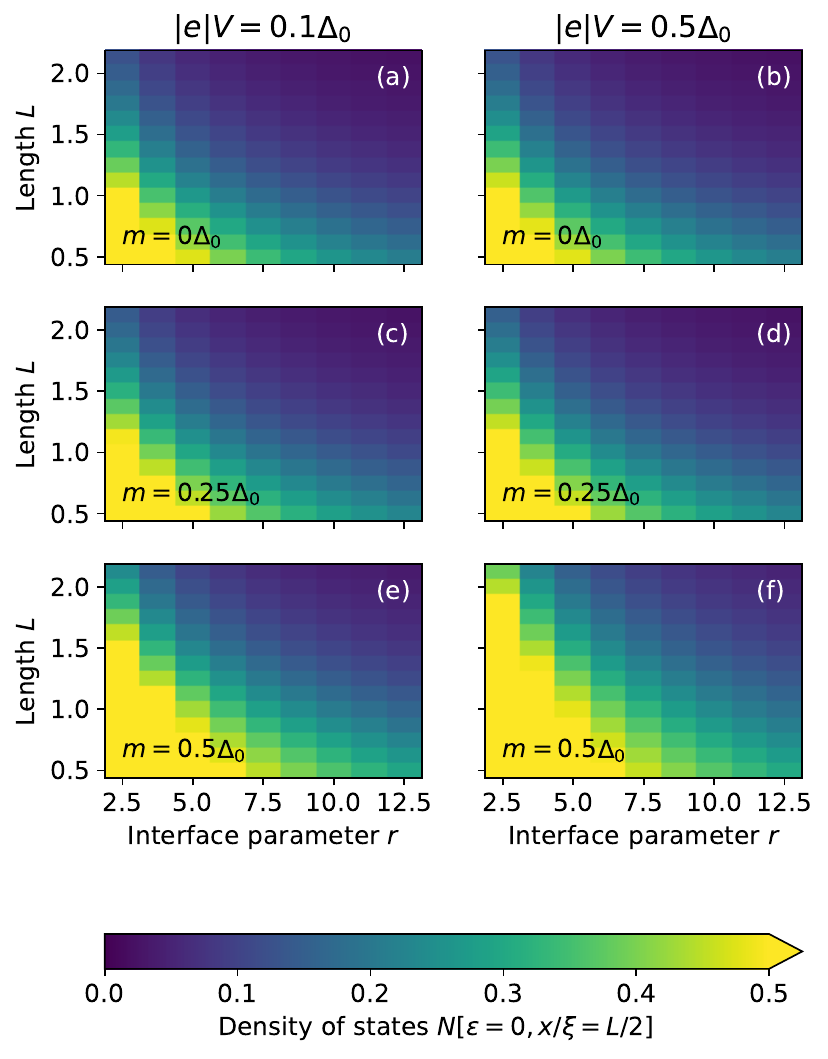}
    \caption{The density of states in the middle of the superconductor at zero energy in the $(r,L)$-plane for different combinations of the spin-splitting $m$ and the voltage bias $V$. }
    \label{fig:DOS-rLplane-standard-voltage}
\end{figure}

% CAR dominates when the proximity effect is small
The strength of the inverse proximity effect inside the superconductor is the crucial factor that regulates the ratio of CAR and EC. We consider the setup shown in Fig. \ref{fig:car-setup}(b), and we calculate the non-local current in the $(r,L)$-plane for two different voltage biases. The results are shown in Fig. \ref{fig:current-rLplane-standard-voltage}(a)-(b). Looking at the blue regions in the plots, we observe that CAR is at its strongest when the superconductor is long and the interface resistance is high. These are parameter sets that decrease the impact of the normal leads. Low transparencies obviously decrease the contact between the normal leads and the superconductor, while a long superconductor is unaffected by the normal leads far from the interfaces. Furthermore, a long superconductor is more favorable for CAR because CAR requires the establishment of a supercurrent. In short superconductors where the supercurrent is small, CAR is suppressed in accordance with the results in Ref. \cite{bergeret_nonlocal_2009}. In the red regions where EC dominates the non-local transport, the superconductor is short or the interface transparency is high. Therefore, the inverse proximity effect in the superconductor is considerable, weakening the superconducting correlations. This is seen in Figs. \ref{fig:DOS-rLplane-standard-voltage}(a)-(b). The zero-energy DOS is high in the regions where EC dominates, while it is smaller in the regions where CAR dominates. The DOS at small energies is important for the transport properties for the following reason. In the grounded regions, electrons occupy the states with $\varepsilon<0$, while they occupy states with $\varepsilon < -|e|V$ in the left normal electrode $N_1$. This is illustrated in Fig. \ref{fig:car-setup}(c). Therefore, low energies $-|e|V\leq\varepsilon\leq0$ are the relevant energies concerning electron transport properties. When the DOS at low energies is high, quasiparticles can flow through the superconductor. This process contributes to EC, and thus EC is strong when the inverse proximity effect is strong. However, when the DOS is sufficiently suppressed at low energies, EC depends on charge imbalance, tunneling through evanescent states, or other processes that do not require available quasiparticle states inside the superconductor. Since CAR requires Cooper pairs it becomes more probable the more intact the superconductor is with respect to the inverse proximity effect, and therefore CAR can surpass EC when the gap suppression is small. We underline that it is not only the influence of the spatially dependent gap on the density of states that is an important consequence of the self-consistency but also the fact that self-consistency correctly describes conversion between resistive and supercurrents, which was not accounted for in previous works studying CAR with quasiclassical theory. As shown in Fig. \ref{fig:iterations}, this is crucial with respect to obtaining the correct result for CAR in non-local transport. 
% Tradeoff: CAR vs good conditions for transport
When we decrease the proximity effect by using less transparent NS contacts or a longer superconductor, we also increase the resistance of the junction. This causes both the local and the non-local currents to decrease. Therefore, we must trade a higher CAR-to-EC ratio against a weaker non-local signal.

\begin{figure}
    \centering
    \includegraphics[width=0.8\linewidth]{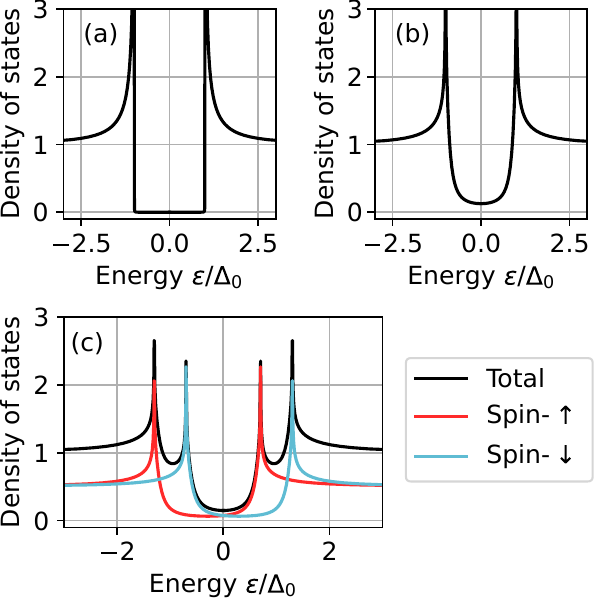}
    \caption{\textbf{(a)} The density of states in a bulk superconductor. \textbf{(b)} The density of states in a superconductor proximitized with a normal metal. \textbf{(c)} The total and spin-resolved density of states in a spin-split superconductor proximitized with a normal metal. The peaks in the density of states now occur at $\epsilon = \pm \Delta_0 \pm m$.}
    \label{fig:DOS}
\end{figure}

% Large voltage biases favor ET
Another limitation on the magnitude of the CAR signal is that the voltage bias must be small enough. When the voltage bias is increased, the region where EC dominates grows. We can see this by comparing Fig. \ref{fig:current-rLplane-standard-voltage}(a) where $|e|V = 0.1\Delta_0$ with Fig. \ref{fig:current-rLplane-standard-voltage}(b) where $|e|V = 0.5\Delta_0$. In a bulk superconductor, the DOS increases sharply as the energy $|\varepsilon|\rightarrow \Delta_0^-$ as seen in Fig. \ref{fig:DOS}(a). Figure \ref{fig:DOS}(b) shows how the proximity effect causes the DOS to increase gradually towards its peaks. Increasing the range of relevant energies by increasing the voltage will therefore increase the number of available states. Consequently, CAR is suppressed relative to EC when the voltage increases.

\begin{figure}
    \centering
    \includegraphics[width=0.7\linewidth]{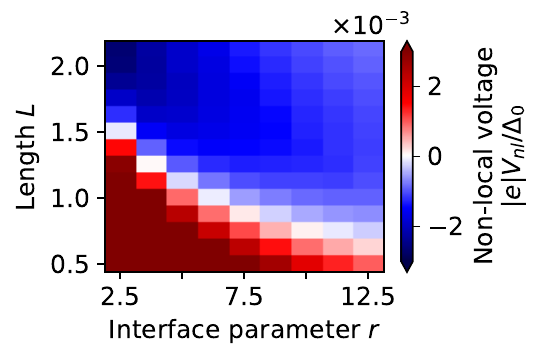}
    \caption{Non-local voltage $|e|V_{nl}/\Delta_0$ in the $(r,L)$-plane for spin-splitting $m=0$ and voltage bias $|e|V/\Delta_0 = 0.1$. This is the same parameter set as used in Fig. \ref{fig:current-rLplane-standard-voltage}(a). A negative non-local voltage (blue in the plot) means that CAR dominates the non-local transport, while a positive non-local voltage (red in the plot) means that EC dominates.}
    \label{fig:induced voltage}
\end{figure}

% Non-local vs non-local voltage bias
Instead of calculating the non-local current, it is possible to calculate the induced non-local voltage in an open circuit setup. Figure \ref{fig:induced voltage} shows the non-local voltage for the same parameters used in Fig. \ref{fig:current-rLplane-standard-voltage}(a). We see that the regions with a positive non-local current $I_ {nl}>0$ are turned into regions with a positive non-local voltage $V_{nl}>0$, and similarly for negative values. Therefore, calculating the voltage or the current gives the same result in terms of whether CAR or EC dominates. This justifies the computationally easier approach of calculating the current when the second lead $N_2$ is grounded. Interestingly, the non-local voltage does not display the decline seen in the non-local current for decreasing interface resistances. Even though the current decreases with increased interface resistance, the non-local voltage required to stabilize the non-local current at zero remains the same. Therefore, it might be experimentally more suitable to measure the non-local voltage as this signal retains its strength even when the leads are weakly connected to the superconductor.

\begin{figure}
    \centering
    \includegraphics[width=\linewidth]{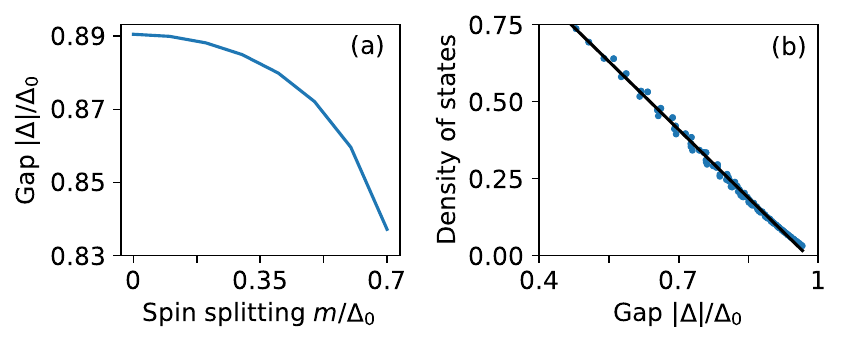}
    \caption{\textbf{(a)} A typical profile of the gap $|\Delta|/\Delta_0$ in the middle of the superconductor $x/\xi=L/2$ as a function of of the spin-splitting field. The gap decreases when the spin-splitting field increases. \textbf{(b)} The relation between the gap and the density of states at zero energy in the middle of the superconductor. The blue markers originate from the combinations of $(r,L)$ in Fig. \ref{fig:current-rLplane-standard-voltage}(a). The black line is the linear regression of the blue markers, revealing a close-to-linear relation between the gap and the zero-energy density of states.}
    \label{fig:spinsplitting decreases DOS}
\end{figure}

% Spin-splitting always favors EC over CAR
Spin-splitting always favors EC over CAR. Fig. \ref{fig:current-rLplane-standard-voltage} shows that the region where EC dominates grows when the spin-splitting increases. This can again be explained by the density of states. We already emphasized that EC dominates when the DOS at low energies becomes large. Spin-splitting always increases the density of states at low energies, and the reason for this is twofold. First, the spin-splitting splits the peaks in the DOS as shown in Fig. \ref{fig:DOS}(c). The proximity effect makes the DOS increase close to the peaks, thus the shifting causes the low-energy DOS to increase. Second, spin-splitting reduces the magnitude of the gap, as demonstrated in Fig. \ref{fig:spinsplitting decreases DOS}(a).  When the magnitude of the gap decreases, the DOS at low energy increases. This is shown in Fig. \ref{fig:spinsplitting decreases DOS}(b). Therefore, CAR cannot be enhanced by spin-splitting. 
% A note on the peaks in DOS
We note that in a previous work \cite{tjernshaugen_superconducting_2024}, the peaks in the DOS were shifted around the actual gap of the superconductor $\varepsilon=|\Delta|$. Now we find that the DOS is shifted around the bulk gap $\varepsilon=\Delta_0$, even when the gap is suppressed far below its bulk value. This is because the superconductor in our model is strongly coupled to a spin-split superconducting reservoir. If the peaks were centered around the actual gap $|\Delta|$, CAR would be further suppressed because the gapped region in the DOS would shrink. 

% Comparison with ballistic system
In ballistic systems, it has been shown that the EC probability always exceeds the CAR probability even when the problem is treated fully self-consistently \cite{melin_self-consistent_2009}. The reason we find a different result in the diffusive limit may be related to the enhancement of AR in diffusive NS structures \cite{wees_excess_1992}. This effect is called reflectionless tunneling. Since the AR probability can be enhanced by impurities in the normal metal, we hypothesize that the CAR probability can be enhanced as well. Verification of this hypothesis would require a separate study, however.

\begin{figure}[t]
    \centering
    \includegraphics[width=\linewidth]{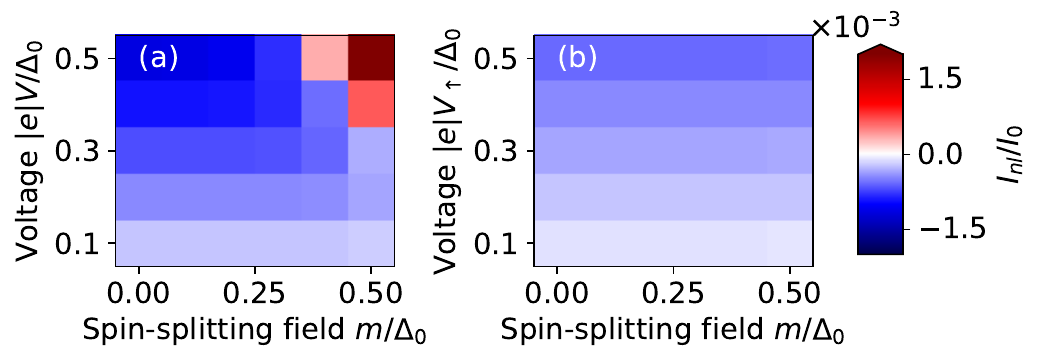}
    \caption{Non-local current $I_{nl}/I_0$ in the $(m,V)$-plane for \textbf{(a)} an electrical voltage, and \textbf{(b)} a spin-up voltage. The superconductor has length $L=1$ and interface resistance $r=10$.}
    \label{fig:mV-plane}
\end{figure}

% Spin dependent transport does not really help CAR
We now turn our attention to the spin-dependent transport scheme shown in Fig. \ref{fig:car-setup}(e)-(f). From the previous discussion, we know that
CAR-dominated transport turns into EC-dominated transport when the spin-splitting and voltage bias increase. This is shown in Fig. \ref{fig:mV-plane}(a). When spin-splitting destroys CAR mainly because the peaks in the DOS are shifted, CAR can be restored by using a spin-up voltage $V_{\uparrow}$. Fig. \ref{fig:mV-plane}(b) shows that the EC-dominated region turns into a CAR-dominated region when the voltage switches from an electrical voltage to a spin-up voltage. 
A prerequisite for this is that CAR must dominate at low spin-splitting fields and low voltages since increasing the voltage bias or the spin-splitting cannot make the spin-resolved DOS smaller at low energies. The reason that the CAR dominance is destroyed in Fig. \ref{fig:mV-plane}(a) is that the spin-down quasiparticle peak in the DOS closes in the low-energy regime. Therefore, EC is strengthened because spin-down electrons can flow through the superconductor. This transport channel is removed by switching to transport of spin-up electrons. For other combinations of $(r,L)$, increasing the electrical voltage at zero spin-splitting causes the probability of EC to exceed that of CAR. We hypothesized that we could restore CAR for high voltages by switching to a spin-up voltage and using spin-splitting. In this case, spin-splitting is necessary because the spin-up voltage gives essentially the same result as the electrical voltage at zero spin-splitting. We discarded this hypothesis since the parameter region in which it is true is very small if it even exists, and thus not particularly interesting. The reason we could not restore CAR by increasing the spin-splitting for high spin-up voltages is that the gap decreases under such conditions, and thus the DOS at low energies increases. Again, we conclude that spin-splitting cannot enhance the probability of CAR.

% The proposed setup for spin dependent transport could find applications elsewhere
Nevertheless, the setup sketched in Fig. \ref{fig:car-setup}(d) demonstrates a new way to achieve spin-dependent transport by tuning the chemical potentials in the leads. Such transport could be interesting for other applications where one needs transport of electrons with one spin-type only. A spin-up voltage also allows for transport of spin-down electrons due to AR or CAR and is thus a different method for achieving spin-dependent transport compared to using spin-polarized leads or interfaces.

\section{Conclusion} 

Summarizing, we have shown that self-consistency is crucial in determining whether crossed Andreev reflection (CAR) or elastic cotunneling (EC) dominates for a system consisting of a superconductor in contact with two normal metal leads. In particular, we show that CAR dominates when the inverse proximity effect weakens, as happens for increasing interface resistance or junction length. We also consider a scenario with spin-splitting in the superconductor, accomplished either via an external in-plane magnetic field for a thin-film superconductor or by growing the superconductor on top of a ferromagnetic insulator. In this case, the spin-splitting increases the subgap density of states and favors EC over CAR. However, when tuning the voltage difference between the leads via spin-pumping so that transport is governed by electrons of one spin type, the CAR probability increases at finite values of the spin-splitting compared to using a purely electric voltage. Our results may be useful as a guide for experiments to select optimal system parameters for the purpose of maximizing CAR and importantly show that even the simplest possible setup with conventional normal metals and a superconductor can provide dominant CAR in a feasible experimental regime. Our results also provide a way to probe non-local transport via pure spin injection into a superconductor.

 \begin{acknowledgments}
B. Brekke is thanked for useful discussions. This work was supported by the Research
Council of Norway through Grant No. 323766 and its Centres
of Excellence funding scheme Grant No. 262633 “QuSpin.” Support from
Sigma2 - the National Infrastructure for High Performance
Computing and Data Storage in Norway, project NN9577K, is acknowledged.
 \end{acknowledgments}

\bibliography{library}

\appendix

\end{document}